\documentclass{article}
\usepackage[T1]{fontenc}
\usepackage[utf8]{inputenc}
\usepackage[]{ismir} 
\usepackage{amsmath,cite,url}
\usepackage{graphicx}
\usepackage{color}

\title{GlobalMood: A cross-cultural benchmark for music emotion recognition}

\multauthor
  {Harin Lee$^{1,2,3}$ \hspace{1cm} Elif Çelen$^1$ \hspace{1cm} Peter Harrison$^4$ \hspace{1cm} Manuel Anglada-Tort$^5$}
  {{\bf Pol van Rijn$^1$ \hspace{1cm} Minsu Park$^6$ \hspace{1cm} Marc Schönwiesner$^3$ \hspace{1cm} Nori Jacoby$^{1,7}$}\\
  $^1$MPI Empirical Aesthetics \hspace{1cm} $^2$MPI for Human Cognitive and Brain Sciences\\
  $^3$Leipzig University \hspace{0.1cm} $^4$University of Cambridge \hspace{0.1cm}
  $^5$Goldsmiths, University of London\\
  $^6$New York University Abu Dhabi \hspace{0.1cm} $^7$Cornell University\\ 
  }

\def\authorname{H. Lee, E. Çelen, P. Harrison, M. Anglada-Tort, P. van Rijn, M. Park, M. Schönwiesner, and N. Jacoby}

\begin{document}

\maketitle

\begin{abstract}
Human annotations of mood in music are essential for music generation and recommender systems. However, existing datasets predominantly focus on Western songs with terms derived from English, which may limit generalizability across diverse linguistic and cultural backgrounds. We introduce `GlobalMood', a novel cross-cultural benchmark dataset comprising 1,180 songs sampled from 59 countries, with large-scale annotations collected from 2,519 individuals across five culturally and linguistically distinct locations: U.S., France, Mexico, S. Korea, and Egypt. Rather than imposing predefined emotion and mood categories, we implement a bottom-up, participant-driven approach to organically elicit culturally specific music-related emotion terms. We then recruit another pool of human participants to collect 988,925 ratings for these culture-specific descriptors. Our analysis confirms the presence of a valence-arousal structure shared across cultures, yet also reveals significant divergences in how certain emotion terms (despite being dictionary equivalents) are perceived cross-culturally. State-of-the-art multimodal models benefit substantially from fine-tuning on our cross-culturally balanced dataset, particularly in non-English contexts. Broadly, our findings inform the ongoing debate on the universality versus cultural specificity of emotional descriptors, and our methodology can contribute to other multimodal and cross-lingual research.



\end{abstract}

\section{Introduction}\label{sec:introduction}
\begin{figure*}[ht!]
  \centering
  \includegraphics[width=\textwidth]{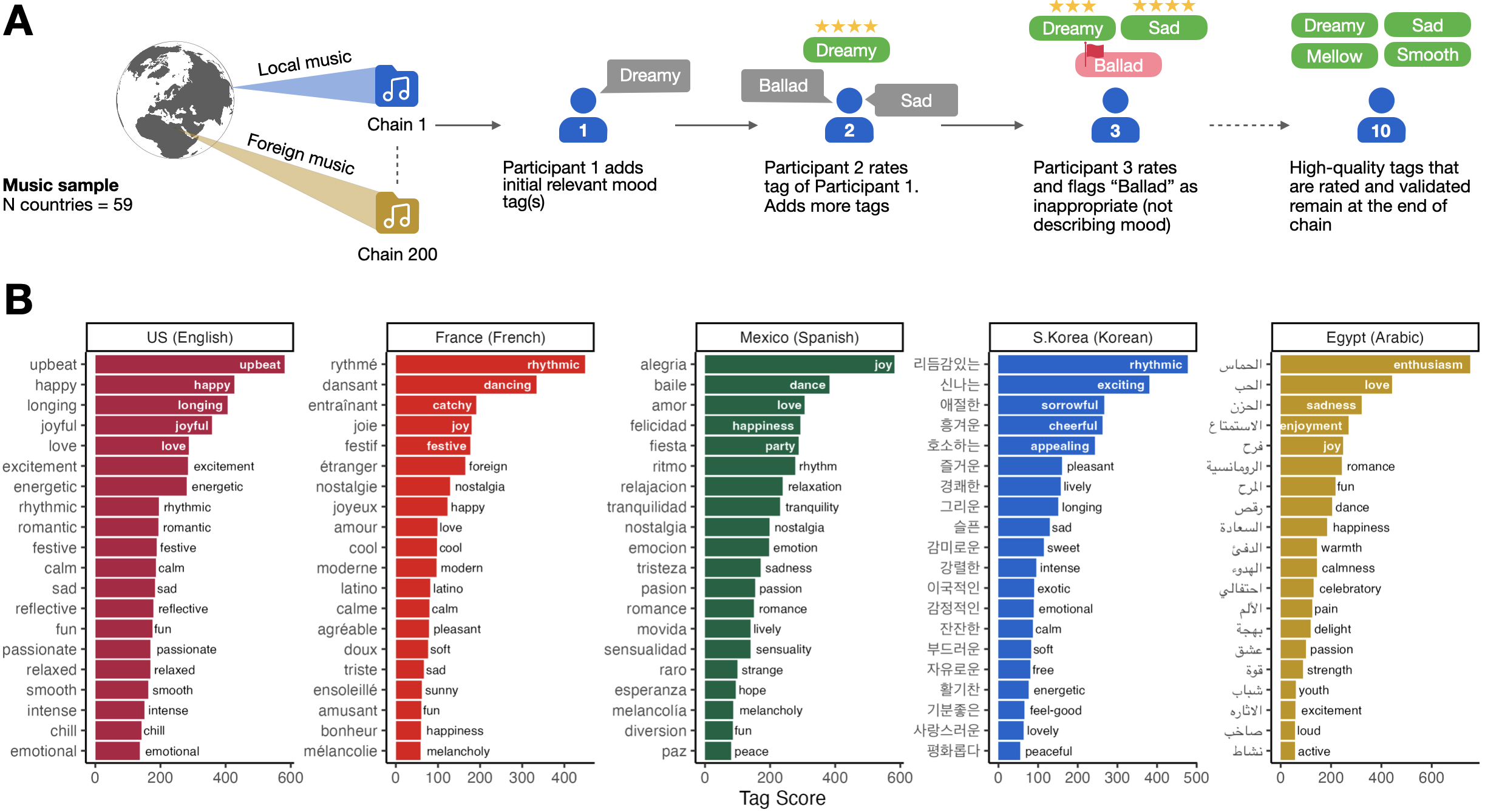}
  \caption{Elicitation and refinement of music emotion terms through iterative participant chains.
  \textbf{(A)} Schematic illustration of the collaborative tagging process within a participant chain. Participants contribute new emotion-related word tags for each song, rate the relevance of existing tags, and can also flag irrelevant content, creating a dynamic refinement system.
  \textbf{(B)} Twenty most reliable emotion tags in each language, ranked by their tag scores. Y-axis labels display tags in their original language (left) and English translations (right).
  }
  \label{fig:top_tags}
\end{figure*}

Music evokes diverse emotional responses in listeners, spanning a wide spectrum beyond basic emotional categories~\cite{zentner2008emotions, cowen2020what}. A central challenge in Music Information Retrieval (MIR) is designing algorithms that can replicate this emotional sensitivity. This is crucial for building recommendation systems that align with listeners' mood and context~\cite{eerola2025what, tran2023emotion, kang2024yetbriefsurveymusic}, and for generating music that resonates with individual preferences~\cite{sun2024emotion}. More broadly, understanding how music conveys emotion is a core question in the science of music~\cite{juslin2008emotional,eerola2012review,park2019global}. To date, however, most algorithms have been trained on datasets derived from Western listeners and Western music, using taxonomies primarily based on English language (e.g., MIREX~\cite{hu2007exploring}).

A significant challenge is creating cross-cultural models capable of handling non-Western music and emotion vocabularies beyond English. Addressing this challenge is essential to developing algorithms that accurately reflect global users' preferences, including those whose musical tastes extend beyond the limited range of styles currently represented in training datasets. Moreover, without capturing culturally specific nuances of emotion, especially those difficult to translate, key aspects of musical meaning may be missed entirely. Direct dictionary translations of English terms may be insufficient, as terms describing emotions are deeply cultural and may lack exact equivalents~\cite{jackson2019emotion,lee2021cross, celen2025expressions, gomez2020joyful}.

To address these issues, we introduce `GlobalMood',\footnote{All code and data: \url{https://github.com/harin-git/GlobalMood}} a new benchmark dataset designed to support culturally inclusive and linguistically diverse emotion and mood recognition in music. Our contribution innovates along three key dimensions: (i) the diversity of musical stimuli, drawn from 59 countries; (ii) the diversity of annotators, spanning five distinct regions (with plans to extens to over 20 languages and locations in future); (iii) a data-driven approach for collecting descriptors, generated organically by participants in their own language during the annotation process. 

Data were collected through two stages involving a total of 2,519 participants and 1,180 songs balanced evenly across 59 countries: In the first stage (Section~\ref{subsec:step}; Figure~\ref{fig:top_tags}), using a smaller subset of 200 songs, we employed our recently developed iterative task that combines open-ended elicitation with collective refinement~\cite{marjieh2023words,van2024giving,celen2025expressions}. Rather than asking listeners to choose from a fixed list of pre-defined emotion terms, we asked them to describe the perceived emotion conveyed in the music using free-text tags in their native language, and at the same time, rate the tags provided by previous listeners. This approach was key to uncovering emotion terms that would otherwise be overlooked by predefined, English-based taxonomies (such as `appeal/plead' that appears in Korean only).

\begin{figure*}[ht!]
  \centering
  \includegraphics[width=\textwidth]{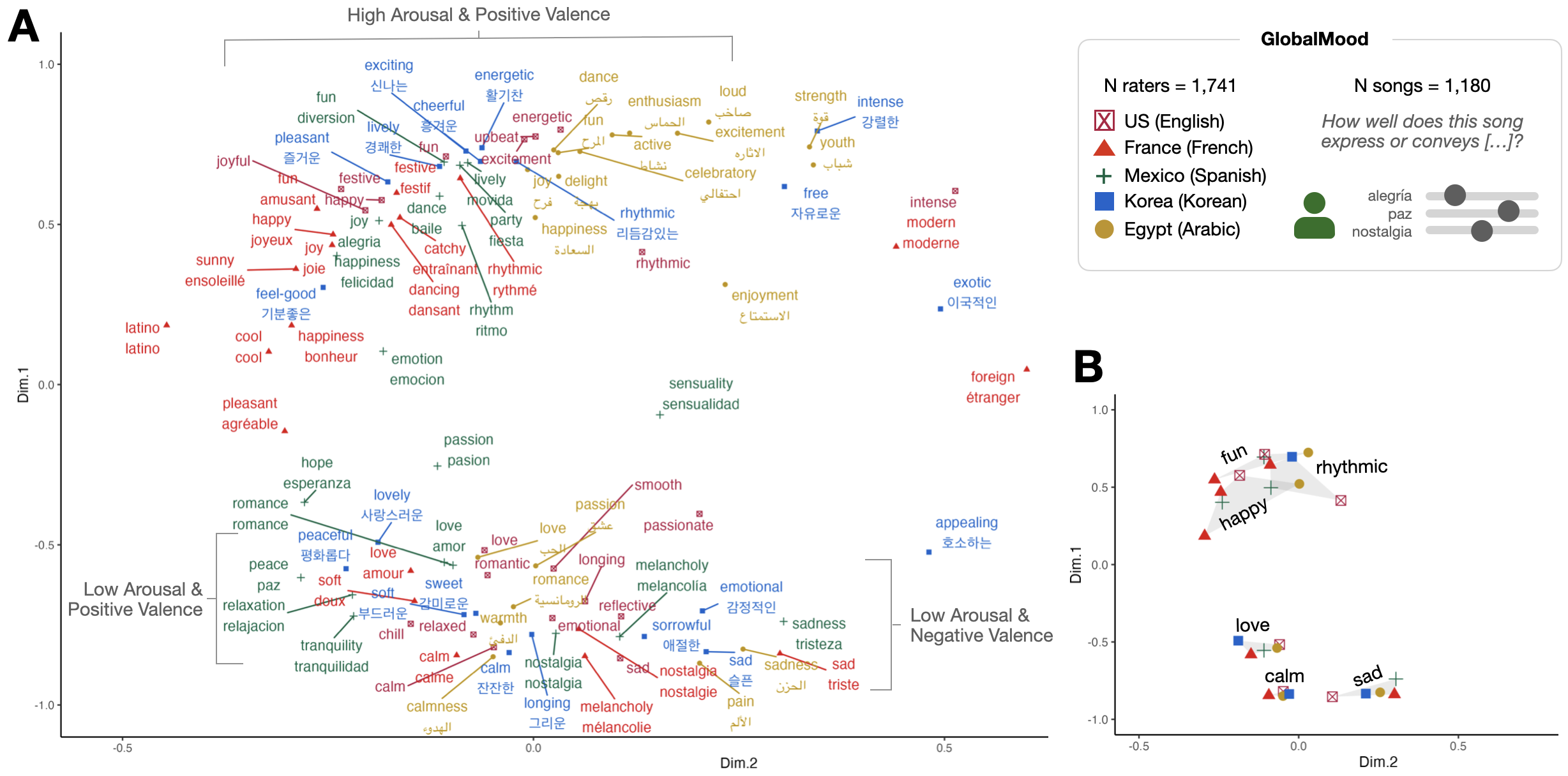}
  \caption{Association between emotion terms across languages.
    \textbf{(A)} MDS visualization of the emotion terms based on mean ratings across the full song set. Terms positioned closer together exhibit similar rating patterns across songs, suggesting similar interpretations across languages.
    \textbf{(B)} Comparison of terms with direct translation equivalents across languages. The area size indicates the degree of semantic divergence despite apparent translation equivalence.
  }
  \label{fig:semantic_map}
\end{figure*}

In the second stage (Section~\ref{subsec:annotation}; Figure~\ref{fig:semantic_map}), we selected the top 20 elicited terms per language and crowdsourced ratings for each tag across the entire set of 1,180 songs. This resulted in a total of 988,925 ratings, creating the most comprehensive open-source cross-cultural emotion annotation dataset in Music Emotion Recognition (MER) to date.

We leveraged GlobalMood to test several recent multimodal and multilingual models (Gemini, CLAP) by evaluating their performance under zero-shot, few-shot, and fine-tuned scenarios (Section~\ref{subsec:multimodal}; Figure~\ref{fig:model_testing}). Models trained only on English data performed poorly in some cultural contexts, but fine-tuning with our cross-cultural data greatly improved their performance in non-English settings. This highlights the critical importance of cross-cultural data in both training MER models and establishing appropriate benchmarks for their evaluation.

\section{Related Works}
\subsection{Music Emotion and Mood Annotation Datasets}
Several datasets have been developed for MER systems with varying annotation approaches.\footnote{Note that databases often extend the concept of emotion to include related constructs such as \textit{mood} or \textit{feeling}. Here we adopt this broader perspective, while acknowledging that subtle distinctions between them do exist.} Early examples include the widely used MIREX 2007 mood dataset~\cite{hu2007exploring} with 240-250 Western songs in five mood clusters derived from AllMusic's English tags (e.g., `passionate–rousing', `wistful–bittersweet'), and CAL500~\cite{turnbull2007towards} with 500 Western pop/rock songs annotated using 18 English mood terms by U.S. undergraduate listeners. Over time, larger datasets appeared: the DEAM corpus (MediaEval `Emotion in Music' dataset~\cite{aljanaki2017developing}) containing 2,058 song excerpts with continuous valence/arousal annotations; mood tags mined from large corpora of Spotify music playlists~\cite{affolter2024utilizing}; and the MTG-Jamendo dataset~\cite{bogdanov2019mtg}, which provides mood/theme tags for 18,486 songs. Notably, Jamendo's tags were freely crowdsourced (56 unique mood labels), which introduced more label variety but still almost entirely in English.

A common limitation across these datasets is their reliance on predefined English descriptors, many of which stem from Western music psychology (for an exception, see Strauss et al.~\cite{strauss2024emotion}). For instance, the Geneva Emotional Music Scale (GEMS) defines 45 emotion descriptors (e.g., `joyful activation') based on studies with European listeners~\cite{zentner2008emotions}, and this taxonomy has been used to annotate datasets like Emotify~\cite{aljanaki2016studying}. Similarly, the mood categories in MIREX and CAL500 were fixed in advance (drawn from AllMusic or prior literature) and presented to annotators as a closed set of options. Consequently, these top-down approaches restrict annotators to the moods the researchers envisioned, leaving any unlisted mood nuances uncaptured and undocumented.

Acknowledging these limitations, recent research has begun exploring MER beyond the Western-centric scope. Hu et al.~\cite{hu2014cross} examined mood annotations of K-pop songs provided by both Korean and American listeners. Their approach involved translating the original MIREX mood categories into Korean for local annotators. Although this allowed direct comparisons of mood classification between Korean and American listeners, it inherently restricted Korean annotations to terms originally defined within Western contexts.

More recently, we compiled a balanced set of American, Brazilian, and Korean songs and gathered mood annotations across nine categories, where annotators rated songs both from their own and the other two countries~\cite{lee2021cross}. We showed that certain mood terms like `energetic' and `sad' are highly consistent across cultures, while more abstract concepts like `love' and `dreamy' diverge considerably. Similar findings have been reported by other studies~\cite{gomez2020joyful, celen2025expressions}, highlighting that when mood descriptors are imposed from one language onto another, important meanings can simply be `lost in translation'.

In summary, while existing MER datasets and research have laid a solid groundwork, they remain limited by insufficient linguistic and cultural diversity. Because many are predominantly English-based and rely on top-down annotation strategies, they may overlook how people in other cultural contexts perceive emotion and mood in music. 



\subsection{Audio LLMs: the New Frontier in Music Tagging}
Recent advances in multimodal large language models (LLMs) have opened promising avenues for downstream MIR tasks, including emotion recognition. These models combine the reasoning capabilities of LLMs with audio perception systems (audio LLMs), enabling more flexible and nuanced music understanding than traditional classification approaches~\cite{li2024mert,weck2024muchomusic,doh2023lpmusiccaps}.


Models like MuLan~\cite{huang2022mulan} and MERT~\cite{li2024mert} have demonstrated potential for zero-shot music emotion and mood classification by embedding audio and natural language descriptions in a shared semantic space. However, comprehensive benchmarks such as the MuChoMusic~\cite{weck2024muchomusic} highlight a crucial limitation: these models rely heavily on language modality and do not attend sufficiently to audio, often failing with more nuanced audio examples for downstream MIR tasks. This limitation could be particularly critical for non-Western music and non-English emotion descriptors, given that their training data are largely from Western contexts.

Similarly, closed-source models (e.g., Gemini) have shown promise in psychological textual analysis in multilingual contexts~\cite{rathje2024-or}, while evaluations in specialized domains such as MIR remain scarce. The proprietary nature of their training data complicates thorough assessment of cross-lingual or cross-cultural performance. We aim to address these fundamental gaps by providing a large set of diverse, multilingual descriptors and annotations to support broader cross-cultural generalizability of audio LLMs.

\section{Method}\label{sec:method}
\subsection{Participants}\label{subsec:participants}
We recruited two independent sets of participants across the two stages of our data collection: \textbf{Stage 1} for emotion term elicitation (N = 778; see Section~\ref{subsec:step}) and \textbf{Stage 2} for subsequent ratings on top 20 terms (N = 1,741; see Section~\ref{subsec:annotation}). Participants had to be at least 18 years old, reside in the target country, and speak the target language as their primary language. Participants from the US were recruited through Prolific, while participants from the other four countries (France, Mexico, S. Korea, and Egypt) were recruited through the CINT platform. All participants provided informed consent under an approved protocol (see Section~\ref{sec:ethics}). Participants were instructed to wear headphones and had to pass a headphone screening task~\cite{milne2021}, and a language proficiency test~\cite{vanrijn2023around} before being eligible for the main experimental task. Experiments were conducted in each participant's native language (English, French, Spanish, Korean, and Egyptian Arabic), with instructions translated using GPT-4o. Code to replicate the experiment through the \textit{PsyNet} framework~\cite{harrison2020gibbs} and all data are available at \url{https://github.com/harin-git/GlobalMood}


\subsection{Globally Representative Song Selection}\label{subsec:stimuli}
To create a globally representative music dataset, we used weekly YouTube top 100 music charts (year 2017-2023) from 59 countries, spanning six continents. To ensure each country's charts reflected its distinct popular music, we excluded any track appearing in more than one country's chart. This left us with a \textit{country-exclusive} pool of songs. From this pool, we sampled 20 songs per country, yielding 1,180 songs in total. This diverse set is designed to capture a wide range of musical traditions and serve as a robust testbed for cross-cultural emotion recognition. Each 15-second audio excerpt was trimmed from a random starting point in the full track, and normalized at -5dB loudness. 

\subsection{Model Evaluation}\label{subsec:models}
We used the resulting GlobalMood dataset to evaluate several recent multimodal and multilingual models capable of music understanding. Specifically, we assessed Google's Gemini models (\textit{1.5 Flash}, \textit{2.0 Flash}, and the latest \textit{2.5 Pro}), a family of multimodal large language models capable of processing and reasoning across text and audio (but also image and video).\footnote{Preliminary tests with other recent multimodal models showed performance issues---Flamingo 2~\cite{ghosh2025audio} struggled with rating consistency and GPT-4o~\cite{hurst2024gpt} failed to generate musical descriptions or ratings from audio alone---thus we excluded them from further analysis.} We compared zero-shot and few-shot approaches, where the latter included 10 human-rated emotion terms as examples.

Given that Gemini is closed-source, we also included CLAP (Contrastive Language-Audio Pretraining)~\cite{wu2023large} as an alternative, open-source model that learns joint audio-text embeddings. CLAP has demonstrated promise in MIR applications~\cite{barnett2024exploring} and serves as the foundation for music-specific models like CLaMP~\cite{wu2025clamp}. Here, we conducted zero-shot evaluations through: (1) extracting audio embeddings from CLAP, (2) computing cosine similarities with text embeddings of emotion terms, and (3) comparing these scores to human ratings.

We also fine-tuned CLAP on GlobalMood (train–test split = 1,000:180) to assess potential performance improvements. To preserve the continuous nature of our ratings, we represented each term in proportion to its mean rating (e.g., the term `calm', with a mean rating of 3.0, appeared three times in the text). This method retained the nuanced information in our soft labels rather than reducing them to binary categories. To improve generalizability, we created 10 augmented variations of each song through pitch shifting (range of $\pm$3 semitones), loudness adjustment (range of $\pm$15dB), and the addition of Gaussian noise (amplitude of 0.005). Each augmented variant randomly included one or two of these modifications.

\section{Results}\label{sec:results}
\subsection{Bottom-up Term Elicitation Across Languages}\label{subsec:step}
\subsubsection{Tagging pipeline}
Many existing studies on music emotions rely on pre-defined taxonomies or web-scraped data that offer limited linguistic diversity~\cite{hu2007exploring, affolter2024utilizing}. To overcome this limitation, we employed a bottom-up, participant-driven tagging method~\cite{marjieh2023words,van2024giving,celen2025expressions}. Specifically, we asked participants in each country to complete independent `chains' of iterative annotations. A subsample of 200 songs from the 1,180 entire set was used as stimuli. This subsample consisted of 180 balanced songs across countries, with an additional 20 local songs drawn from the participating country's pool. This was to ensure that local participants encounter enough music strongly tied to their background, allowing them to elicit culturally specific emotion descriptors.

Figure~\ref{fig:top_tags}A illustrates one such chain: (i) The first participant annotates the song using single-word emotion tags in their native language; (ii) The second participant (from the same country) rates the relevance of these tags (1--5 scale), flag irrelevant tags (e.g., genre- or lyrics-related rather than emotion), add new tags as necessary; (iii) The third participant sees all tags from earlier participants and repeats these steps; (iv) This iterative process continues through ten participants per chain, systematically refining and validating emotion terms. In each country, we ran the entire elicitation experiment twice and aggregated the results to increase the diversity of responses from a larger pool of participants.

\subsubsection{Top emerging terms}
Following the removal of tags flagged by more than two participants in a chain, our STEP-Tag process yielded an extensive, culturally specific lexicon of emotion terms across languages (N unique terms: English = 644; French = 528; Spanish = 870; Korean = 629; Arabic = 283). To identify the most salient terms in each language, we calculated a composite score for every term by multiplying its frequency of occurrences across chains by its mean relevance rating. Higher scores indicate terms frequently mentioned and consistently rated as highly relevant.

We consolidated closely related morphological variants (e.g., `happy' and `happiness' in English; gendered forms such as `joyeux' and `joyeuse' in French) manually with native speakers. Figure~\ref{fig:top_tags}B presents the resulting 20 highest-ranking tags per language, displaying both the original word and English translations to facilitate cross-cultural comparisons. 

Despite being explicitly asked to provide emotion terms, participants often gave broader affective descriptors like moods or feelings (e.g., `soft’ and `festive’). This aligns with prior MIR literature that often includes both emotion and mood, and given its relevance for practical use, we did not enforce a strict distinction.

\subsection{Large-scale Diverse Human Ratings}\label{subsec:annotation}
\subsubsection{Cross-cultural ratings across the entire set}
Having identified the top 20 terms for each language, we next gathered exhaustive ratings for the entire 1,180 songs of GlobalMood. We recruited 1,741 new participants (see Section~\ref{subsec:participants}) who listened to the 15-second excerpts and rated how effectively each excerpt conveyed a given emotion and mood term (1--5 scale). For each stimulus, participants evaluated seven randomly selected terms from the relevant language set. This systematic approach ensured that, on average, each song in each language received 8.38 (SD = 2.40) unique participant ratings, resulting in an extensive collection of 988,925 ratings spanning across five languages.

\subsubsection{Is `happy' in my language the same `happy' in your language?}
To investigate differences in how each culture interprets these terms, we constructed 100 rating vectors (5 languages $\times$ 20 terms per language). Each vector was 1,180-dimensional, capturing the mean rating per term across the 1,180 song set. We then performed non-metric multidimensional scaling (MDS) using correlation as the distance metric, projecting these vectors in a two-dimensional space. In this emotion `space,' terms that position close to one another---even those from different languages---reflect similar rating patterns across the musical examples, suggesting comparable emotional interpretations across cultures.

Figure~\ref{fig:semantic_map}A visualizes this emotion space. The terms cluster into two main regions: one region of high arousal and high valence (e.g., \textit{happy}, \textit{energetic}, and \textit{lively}; upper region of the figure) and a second region of low arousal that spans positive valence (e.g., \textit{peaceful}; bottom left) to negative (e.g., \textit{sad}; bottom right). Notably, many translated `equivalents' appear close together, which might suggest a general cross-cultural consensus on what music evokes what emotions.

However, examining six commonly shared terms that have direct translations in at least four of the five languages (\textit{fun}, \textit{happy}, \textit{rhythmic}, \textit{love}, \textit{sad}, and \textit{calm}) revealed varying degrees of cross-cultural agreement (see Figure~\ref{fig:semantic_map}B). For each of these terms, between-country agreement ($r_{\text{between}}$) was computed as the average of pairwise correlation coefficients, while within-country agreement ($r_{\text{within}}$) was calculated using split-half reliability with Spearman-Brown formula. Effectively, $r_{\text{within}}$ serves as measurement error to compare as baselines when evaluating $r_{\text{between}}$.  

The term \textit{calm} showed the highest average agreement ($r_{\text{between}}$ = 0.52 [0.49, 0.55]; $r_{\text{within}}$ = 0.49 [0.38, 0.57]), followed by \textit{fun} ($r_{\text{between}}$ = 0.46 [0.39, 0.53]; $r_{\text{within}}$ = 0.44 [0.27, 0.54]), \textit{love} ($r_{\text{between}}$ = 0.44 [0.41, 0.47]; $r_{\text{within}}$ = 0.47 [0.33, 0.66]), \textit{sad} ($r_{\text{between}}$ = 0.41 [0.37, 0.45]; $r_{\text{within}}$ = 0.43 [0.32, 0.58]), \textit{rhythmic} ($r_{\text{between}}$ = 0.38 [0.30, 0.45]; $r_{\text{within}}$ = 0.43 [0.25, 0.53]), and notably \textit{happy} ($r_{\text{between}}$ = 0.37 [0.28, 0.45]; $r_{\text{within}}$ = 0.45 [0.24, 0.59]). 

Overall, considering within-country agreement (mean $r_{\text{within}}$ = 0.43--0.48), most of these terms were comparable in their between-country agreement (mean $r_{\text{between}}$ = 0.39--0.52). However, despite being considered a basic universal human emotion~\cite{ekman1992basic}, \textit{happy} exhibited a considerable gap between between- and within-country agreement. This emphasizes the necessity of incorporating diverse cultural perspectives when modeling nuanced musical emotional responses. Reliance on either dictionary translation or LLM-based translation alone could overlook important, context-specific nuances in emotion and mood perception---particularly relevant when building models for global audiences.




\begin{figure}[ht!]
  \center
  \includegraphics[width=\columnwidth]{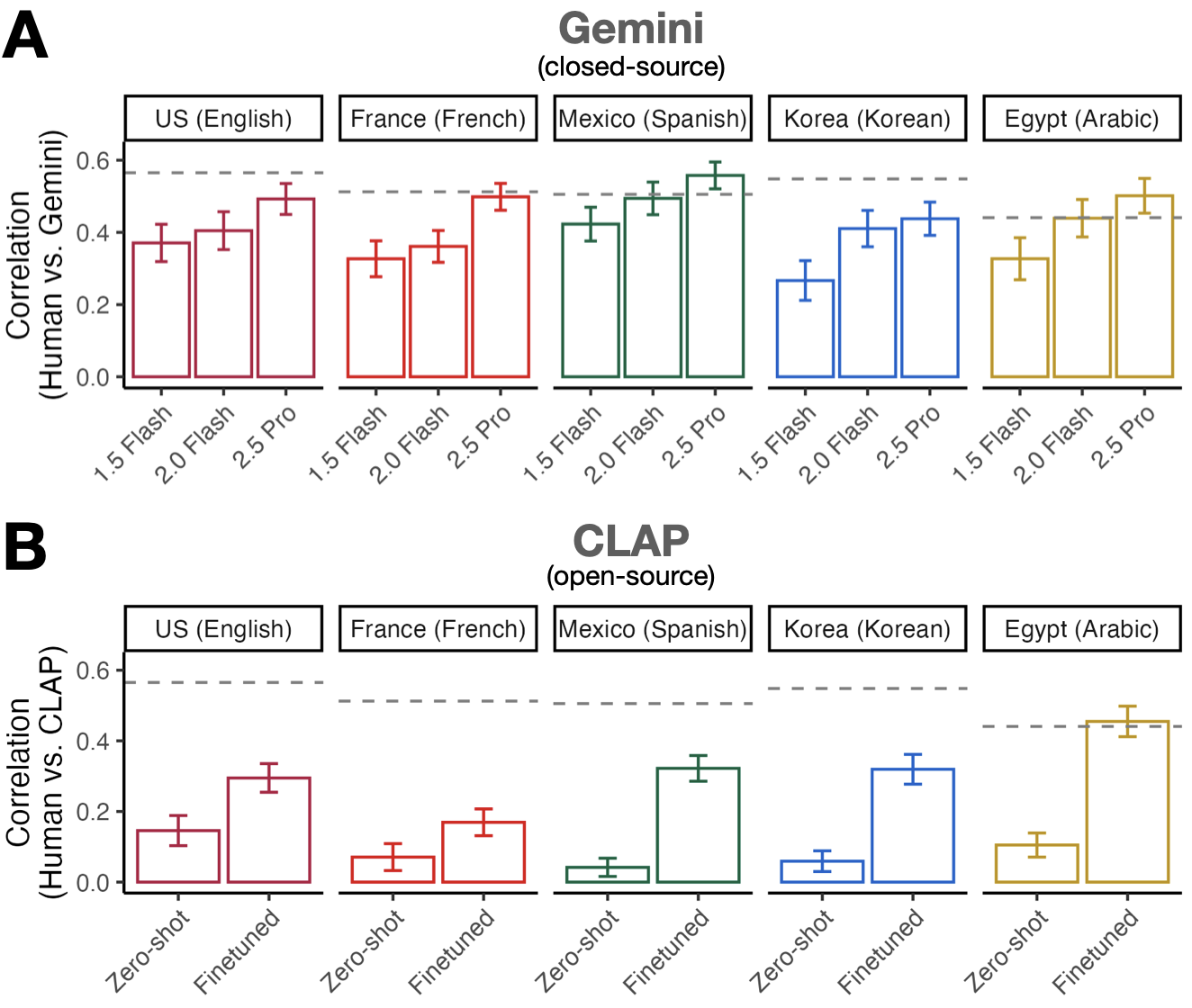}
  \caption{Correlations between human ratings and multimodal model predictions. \textbf{(A)} Gemini models with zero-shot prompting showing increase in performance with newer models. \textbf{(B)} CLAP models in zero-shot and fine-tuned scenarios showing how the use of multilingual annotations can substantially increase performance. Gray dashed lines represent split-half reliability of human ratings using the Spearman-Brown formula as baseline reference of correlations achieved between humans. Error bars indicate 95\% CI of mean correlation across songs.
  }
  \label{fig:model_testing}
\end{figure}

\subsection{Human vs. Multimodal Models}\label{subsec:multimodal}
Recent benchmarks have evaluated the capabilities of audio LLMs across various downstream MIR tasks, but these evaluations have also been restricted to English~\cite{weck2024muchomusic}. We evaluated both closed-source (Gemini) and open-source (CLAP) models against our GlobalMood multilingual human ratings (see Section~\ref{subsec:models} for model details).

For Gemini (Figure~\ref{fig:model_testing}A), we replicated our human study protocol by prompting the model with `Rate from a scale of 1 to 5 how well this song expresses or conveys the emotion [...]' independently for the five native languages. We then evaluated how well the model's outputs predict human judgments. We observed consistent improvement with each new model version. The earliest model, \textit{1.5 Flash}, demonstrated modest alignment with human judgments (mean correlation across countries: $r$ = 0.34 (95\% CI = [0.27, 0.42]). This model struggled particularly with Korean language ratings ($r$ = 0.27 [0.21, 0.33]). The subsequent \textit{2.0 Flash} version bridged this gap in Korean ($r$ = 0.42 [0.36, 0.47]), and achieved a substantially higher mean correlation of $r$ = 0.42 [0.36, 0.48]. The latest \textit{2.5 Pro} model demonstrated another leap with a mean correlation of $r$ = 0.50 [0.45, 0.55]. Few-shot approach with 10 human-rated examples using 2.5 Pro did not improve the results ($r$ = 0.47 [0.42, 0.52]). This consistent upward trajectory across model iterations provides compelling evidence that these general-purpose systems are progressively developing more sophisticated understanding of musical emotions across diverse linguistic and cultural contexts.

This level of correlations between Gemini and human ratings are on par with algorithms specifically designed for MER, such as Spotify's mood estimation~\cite{lee2021cross, Duman2022-jd}. Moreover, given the subjective nature of emotion and mood perception in music (where even humans often disagree), the latest Gemini model already reaches human-level performance, matching the theoretical upper bound defined by inter-rater human agreement (gray dashed lines in Figure~\ref{fig:model_testing}).

For CLAP, an open-source alternative (Figure~\ref{fig:model_testing}B), we evaluated both a zero-shot approach and a fine-tuned version trained on our GlobalMood dataset (see Section~\ref{subsec:models} for fine-tuning details). The zero-shot approach measured the similarity between the emotion term's text embedding and the song's audio embedding. This zero-shot CLAP performed poorly (mean $r$ = 0.08 [0.03, 0.13]), while fine-tuning with GlobalMood substantially improved the performance (mean $r$ = 0.31 [0.19, 0.44]). As a control experiment, we also fine-tuned CLAP using a dataset where all non-English emotion terms were first translated into English using an LLM without any musical context (instead of original emotion terms collected from native speakers). This translation-based approach failed to improve performance (mean $r = 0.13$ [-0.05, 0.32]), showing that the performance gains from fine-tuning come from culturally-specific information rather than from mere increase in data volume. 

Importantly, improvements through finetuning were most pronounced for non-English languages. Based on Fisher's z test for correlation comparisons, Arabic showed the largest increase from $r$ = 0.11 to 0.46 ($z$ = 0.39), followed by Spanish (from 0.04 to 0.32; $z$ = 0.29) and Korean (from 0.06 to 0.32; $z$ = 0.27). The least substantial increase was observed for French (from 0.07 to 0.17; $z$ = 0.10). These observations demonstrate the promising potential of how native-language terms and ratings from our GlobalMood dataset can be used to generalize to new cultures and languages, when modeling MER systems.

\section{Discussion}\label{sec:discussion}

We introduce GlobalMood, a novel cross-cultural benchmark dataset comprising emotion and mood descriptors and ratings collected from a large and globally diverse participant pool through a data-driven approach. Consistent with previous research in music consumption and perception~\cite{jacoby2019universal,mehr2019universality,jacoby2024commonality,savage2015statistical,mcdermott2016indifference,Lee2025-ft} and music emotion~\cite{lee2021cross,cowen2020music,celen2025expressions,gomez2020joyful,fritz2009universal}, our findings highlight both cross-cultural similarities and differences. Specifically, we demonstrate that music emotion descriptors across languages are broadly organized around clusters relating to valence and arousal~\cite{eerola2012review,ekman1992basic}.

Relying on dictionary translations for music emotion terms may face limitations (e.g., \cite{cowen2020music}), where cross-lingual agreement varied across terms. For instance, despite \textit{happy} being a basic emotion, it exhibited a relatively large gap between cross- and within-country agreement. This highlights the need for incorporating multilingual descriptors from diverse annotators. Our future work aims to expand GlobalMood to reach over 20 languages~\cite{vanrijn2023around,niedermann2023studying} and a broader range of musical styles, including works from different historical periods, which will be critical for developing more robust tools that can generalize across diverse cultural contexts.


Notably, fine-tuning CLAP on cross-cultural data significantly boosted performance in non-English contexts, highlighting the value of GlobalMood. These findings point to the potential for culturally-sensitive MER systems, moving beyond one-size-fits-all models. Future work should investigate more varied prompting and data augmentation techniques, and alternative LLM architectures.

Our tagging pipeline is quick to deploy across cultures and uses a bottom-up approach that helps minimize researcher bias. This offers an advantage over predefined term lists, which often fail to capture the cultural subtleties of emotion terms~\cite{blasi2022over}. However, it has drawbacks: early participants can bias the tag pool and there’s no theoretical guarantee that all tags reflect clear emotion concepts (e.g., `étranger' in French). Unlike structured systems for crowdsourcing tags (e.g., TagATune~\cite{Law2007TagATune}), our approach trades control for flexibility and cultural responsiveness.


Beyond MIR, the GlobalMood and associated pipeline offer timely implications for broader multimodal and cross-lingual research, particularly in Natural Language Processing (NLP) communities~\cite{hershcovich-etal-2022-challenges}. As NLP increasingly tackles multimodal tasks involving audio–text modeling and cross-cultural applications, our dataset may provide a useful benchmark for evaluating language–audio models across diverse contexts. The iterative annotation pipeline can also serve as an effective framework for collecting representative samples and annotations in other domains \cite{van2024giving,huang2024characterizing,marjieh2023words,harrison2020gibbs}.

\section{Acknowledgments}
H.L. was funded by the Max Planck Society. M.P. was partially supported by the NYUAD Center for Interacting Urban Networks (CITIES), funded by Tamkeen under the NYUAD Research Institute Award CG001.

\section{Ethics Statement}\label{sec:ethics}
We conducted our human experiment according to ethical best practices. All participants recruited via Prolific or CINT provided informed consent based on an approved protocol (Max Planck Ethics Council \#202142). Participant data was collected anonymously (except for Prolific or CINT IDs), and all published data are fully anonymized. Models were accessed through commercial APIs, with fine-tuning performed locally. Conducting cross-cultural research requires sensitivity to diverse ethical considerations \cite{jacoby2020cross}, and this study was designed accordingly. We acknowledge that machine learning models and training datasets may contain biases originating from participants, data selection processes, and the models themselves. Despite these potential biases, our study aims to address critical limitations and societal risks within current MIR technologies, which predominantly focus on English-language music and raters despite serving diverse global populations \cite{henrich2010most,blasi2022over,amiri2024beyond}.

\bibliography{references}

\end{document}